**Highly enhanced Curie temperatures in low temperature annealed (Ga,Mn)As epilayers**

K. C. Ku,* S. J. Potashnik,* R. F. Wang,* M. J. Seong, † E. Johnston-Halperin,# R. C. Meyers, # S. H. Chun,* A. Mascarenhas, † A. C. Gossard,# D. D. Awschalom,# P. Schiffer,* and N. Samarth*[(a)]

**Department of Physics and Materials Research Institute, Pennsylvania State University, University Park, PA 16802, USA.*

†*National Renewable Energy Laboratory, Golden CO 80401, USA.*

#*Center for Spintronics and Quantum Computation, University of California, Santa Barbara, CA 93106, USA.*

ABSTRACT

We report Curie temperatures up to 150 K in annealed $Ga_{1-x}Mn_xAs$ epilayers grown with a relatively low As:Ga beam equivalent pressure ratio. A variety of measurements (magnetization, Hall effect, magnetic circular dichroism and Raman scattering) show that the higher ferromagnetic transition temperature results from an enhanced free hole density. The data also indicate that, in addition to the carrier concentration, the sample thickness limits the maximum attainable Curie temperature in this material – suggesting that the free surface of $Ga_{1-x}Mn_xAs$ epilayers is important in determining their physical properties.

[(a)] nsamarth@psu.edu

The ferromagnetic semiconductor $Ga_{1-x}Mn_xAs$ – first developed using low temperature molecular beam epitaxy (MBE) in 1996 [1,2] – continues to elicit substantial interest for semiconductor spintronics.[3,4] Although a detailed understanding of hole-mediated ferromagnetism in this material still remains the subject of active discussion, mean field approaches[5,6] suggest that the Curie temperature ($T_C$) increases with the free hole density ($p$) and could reach values above 300 K if $p$ were sufficiently large. Hole-compensation by defects has thus far limited $T_C$ to a maximum value of ~110 K in optimally as-grown epilayers (x ~ 0.05)[7] as well as in post-growth annealed[8] samples ($0.05 < x < 0.083$).[9,10] Such observations prompted the suggestion that $T_C$ ~ 110 K could be limited by fundamental constraints on the maximum attainable free hole concentration in low temperature grown $Ga_{1-x}Mn_xAs$.[11] Very recent work, however, has revived promise of higher temperature ferromagnetism in $Ga_{1-x}Mn_xAs$ by the post-growth annealing of a variety of $Ga_{1-x}Mn_xAs$ samples that include modulation-doped digital alloy heterostructures ($T_C \approx$ 170 K),[12] epilayers co-doped with Be acceptors ($T_C \approx$ 150 K),[13] and "standard" $Ga_{1-x}Mn_xAs$ epilayers ($T_C \approx$ 140 K).[14]

Here, we demonstrate an alternative combination of growth and annealing parameters that consistently yields $Ga_{1-x}Mn_xAs$ epilayers with 110 K $\leq T_C \leq$ 150 K. These results are reproduced in two different MBE systems using either $As_2$ or $As_4$. We find that *both* the carrier concentration and the epilayer thickness influence the maximum attainable $T_C$, which has strong implications for heterostructure design. Furthermore, measurements of $T_C$ vs. $p$ over an order of magnitude in hole density elucidate the prospects for achieving room temperature ferromagnetism in $Ga_{1-x}Mn_xAs$.

In order to establish the general applicability of the growth protocol described here, the $Ga_{1-x}Mn_xAs$ samples are grown in two different MBE systems (EPI 930 and Varian GenII), that use uncracked $As_4$ and cracked $As_2$ ($T_{cracker}$ ~ 690° C) respectively for the arsenic source. Another difference between the two MBE systems lies in the measurement of the substrate temperature: the EPI system uses a radiatively coupled thermocouple to measure the temperature of substrates that are In-bonded to a molybdenum block, while in the Varian system, an optical monitoring system allows direct measurement and control of substrates held in an In-free mount.[15] In this paper, we principally focus on samples grown in the former system, with samples grown in the latter confirming the principal results. All samples are grown on (001) semi-insulating, epiready GaAs substrates using growth conditions that are similar to those described elsewhere[9] except for a lower As:Ga beam equivalent pressure ratio (15:1 as compared with the 20:1 ratio used in the earlier work). $Ga_{1-x}Mn_xAs$ epilayers with thickness ($t$) in the range $10 < t < 100$ nm are deposited on a buffer structure that consists of a high temperature GaAs epilayer grown under standard conditions, followed by a low temperature GaAs epilayer grown at 250 ºC. A clear (1x2) reconstruction is observed in the reflection high energy electron diffraction during growth of the $Ga_{1-x}Mn_xAs$ epilayers. The as-grown wafers are subjected to post-growth annealing for 90 minutes at 250 ºC in a nitrogen atmosphere.

Both as-grown and annealed samples are characterized using a battery of techniques whose full details are provided elsewhere:[9,16,17] these include electron microprobe analysis (EMPA), x-ray diffraction (XRD), superconducting quantum interference device (SQUID) magnetometry, magneto-resistance, magnetic circular dichroism (MCD), Hall effect, and Raman scattering. The last two measurements indicate that the higher $T_C$ $Ga_{1-x}Mn_xAs$ samples are indeed random alloy ferromagnetic semiconductors and not a mixture of $Ga_{1-x}Mn_xAs$ with metallic ferromagnetic MnAs nanoclusters.

Figure 1 (a) shows magnetization data as a function of temperature for 2 samples of $Ga_{0.915}Mn_{0.085}As$ with $t$ = 15 nm and 50 nm, both as grown and after annealing. The data demonstrate that $T_C$ is increased above 110 K after annealing in both samples, reaching a maximum value of $T_C \sim 150$ K for the 15 nm thick sample. Figure 1(b) shows the thickness dependence of $T_C$ (discussed in detail below), with all other growth parameters kept nominally identical. Figures 2(a) and 2(b) demonstrate the enhanced $T_c$ in these samples through hysteresis loops and Hall effect measurements, and also show that the qualitative properties of these samples are similar to those reported previously for $Ga_{1-x}Mn_xAs$. The coercive fields ($H_C < 20$ Oe even at temperatures down to 5 K) are consistent with the typical values found in the most metallic random alloy $Ga_{1-x}Mn_xAs$ samples in earlier studies.[18] The temperature dependence of the spectrally-resolved MCD signal also confirms the ordering temperature obtained from magnetization data (Fig. 2(c)). However, while the amplitude of the MCD varies with the magnetization as a function of both temperature and field, the spectrum does not reveal an absorption edge between 600 – 900 nm, consistent with metallic behavior.

In order to quantitatively examine the relationship between $T_C$ and the hole density $p$, we use Raman scattering to correlate the hole density with the relative integrated intensities of the coupled plasmon-longitudinal optical phonon mode (CPLOM) and the unscreened longitudinal optical phonon mode (ULO).[16] Representative data from such a measurement in a 50 nm $Ga_{1-x}Mn_xAs$ sample are shown in Fig. 3(a), and the carrier density can be reliably extracted by deconvoluting the Raman lineshape into the CPLOM and ULO components.[16] The analysis shows that the high $T_C$ samples have a very high carrier density (~1.5 x $10^{21}$ $cm^{-3}$), consistent with the spectral dependence of the MCD. We can place these results within a larger context by examining the variation of $T_C$ with hole density $p$ for a wide range of annealed samples (Fig. 3(b)) from our previous studies.[10] While these data are limited to the range of sample preparation techniques we used, an extrapolation indicates that enhancing the $T_C$ of $Ga_{1-x}Mn_xAs$ to 300 K would require an additional order of magnitude increase in $p$.

It is clear from this work and previous studies [8,9,10] that post-growth annealing is required to maximize the $T_C$ of $Ga_{1-x}Mn_xAs$. While the role of annealing is still unclear, recent ion channeling studies[11] have given some insights to the problem. These experiments show that a significant fraction of Mn ions in $Ga_{1-x}Mn_xAs$ occupies non-substitutional, interstitial sites ($Mn_I$), and that annealing decreases the concentration of $Mn_I$. Since $Mn_I$ acts as a donor and hence compensates holes, optimal annealing increases $p$ and correspondingly enhances $T_C$. The enhanced hole density in our annealed samples is consistent with such a scenario, although the relatively small increase in the ferromagnetic moment in some of our samples implies that a large fraction of Mn ions are still not contributing to the ferromagnetic state even after annealing.[10]

In addition to the effects of annealing, an important clue to the limitations on $T_C$ in $Ga_{1-x}Mn_xAs$ comes from the thickness dependence of $T_C$ in our samples, shown in Figure 1 (b) for a consistent set of $Ga_{0.915}Mn_{0.085}As$ samples wherein all the other growth parameters are nominally identical. The data indicate that – for the present set of growth and annealing conditions – the highest $T_C$ in both as-grown and annealed samples occurs for sample thickness $t$ between 10 nm and 50 nm and that $T_C$ is suppressed for larger $t$. The increase in $T_C$ in thinner samples appears to be a general phenomenon,[19] and we have not succeeded in achieving $T_C > 110$ K for annealed samples with $t > 50$ nm. Preliminary experiments carried out in the Varian system (wherein the substrate temperature is directly controlled during growth) suggest that the thickness effect is unrelated to changes in substrate temperature during growth. While it is difficult to speculate in detail without additional microscopic information, it seems that the thickness dependence of $T_C$ may be associated with the proximity of a free surface. Indeed, a recent theoretical calculation suggests that the Mn ions have rather different energetics near a free surface.[20] Furthermore, we find that $Ga_{1-x}Mn_xAs$ epilayers capped with a thin epilayer (~10 - 50 nm) of GaAs typically show a *decrease* in $T_C$ upon annealing, rather than an increase, indicating that the nature of the surface of the $Ga_{1-x}Mn_xAs$ directly effects the properties of the entire epilayer.

In summary, we have shown that $Ga_{1-x}Mn_xAs$ epilayers can be prepared with Curie temperatures as high as 150 K, and that the thickness of the epilayers plays an important role in determining the maximum obtainable $T_C$. These results are important for possible device applications incorporating $Ga_{1-x}Mn_xAs$, since extrapolation of our data implies that room temperature ferromagnetism in $Ga_{1-x}Mn_xAs$ would need almost an additional order of magnitude increase in the free hole density. Perhaps more importantly, the physical properties of $Ga_{1-x}Mn_xAs$ apparently depend on both the thickness of the layers and on whether the $Ga_{1-x}Mn_xAs$ has a free surface. This indicates that caution must be exercised in extrapolating the properties of $Ga_{1-x}Mn_xAs$ measured in isolated epilayers to those of epilayers incorporated within more complex heterostructures.

This research has been supported by ONR N00014-99-1-0071, ONR N00014-99-1-0077, ONR N00014-99-1-0716, DARPA/ONR N00014-99-1093, DARPA/ONR N00014-99-1096, DARPA N00014-00-1-0951, and NSF DMR 01-01318.

**Figure Captions**

Figure 1: (a) Magnetization (in units of Bohr magnetons/Mn) vs. temperature for $Ga_{1-x}Mn_xAs$ (x ~ 0.085) epilayers with $t$ = 15 nm and 50 nm, as grown and after annealing. (b) Curie temperature for $Ga_{1-x}Mn_xAs$ (x ~ 0.085) epilayers of varying thickness.

Figure 2: (a) Magnetization hysteresis loops shown at different temperatures for the 15 nm thick sample in Fig. 1 ($T_C$ ~ 150 K). All measurements are performed with the field in-plane.
(b) Hall resistance vs. magnetic field at different temperatures for the 15 nm sample in Fig. 1 ($T_C$ ~ 150 K). All measurements are performed with the field perpendicular to the sample plane. (c) Temperature dependence of the amplitude of the MCD signal measured at 825 nm and 690 nm. The data are taken in transmission geometry using a 50 nm $Ga_{1-x}Mn_xAs$ (x ~ 0.085) epilayer grown on a stop etch layer of $Ga_{0.55}Al_{0.45}As$.

Figure 3: (a) Raman spectra at 300 K of a 50 nm thick $Ga_{1-x}Mn_xAs$ (x = 0.085) epilayer before and after annealing. The spectra are measured with 457 nm excitation in the z(Y,Y)z configuration. (b) Curie temperature (from magnetization) vs. hole density (from Raman scattering) for a wide range of post-growth annealed $Ga_{1-x}Mn_xAs$ samples of different Mn content. The samples with $T_C$ □ 110 K are all 100 nm thick, while the two samples shown with $T_C$ ~ 140 K are 50 nm thick.

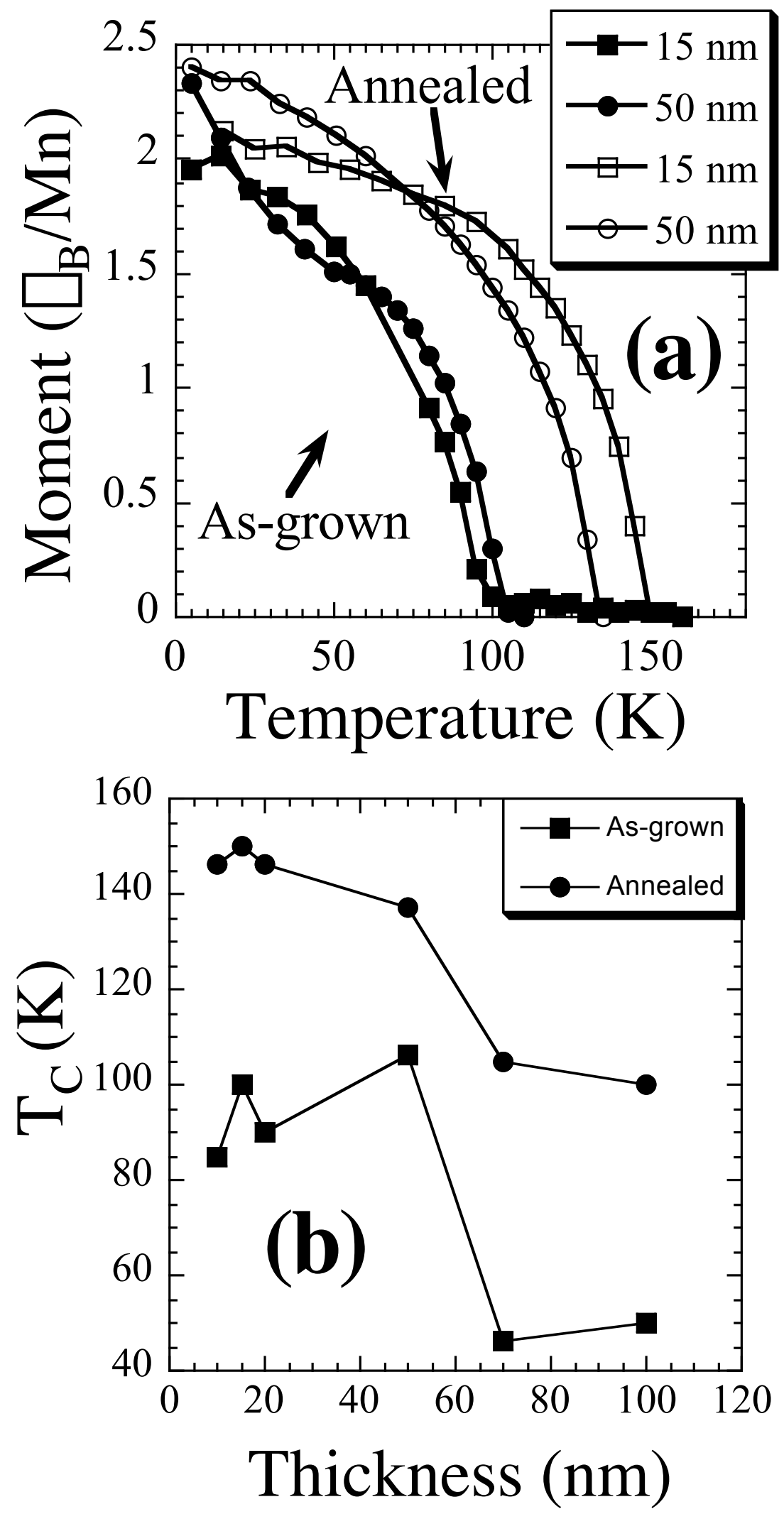

15 nm
50 nm
15 nm
50 nm
Annealed
As-grown
(a)
Moment (□B/Mn)
Temperature (K)
As-grown
Annealed
(b)
TC (K)
Thickness (nm)

Ku *et al* Fig. 1

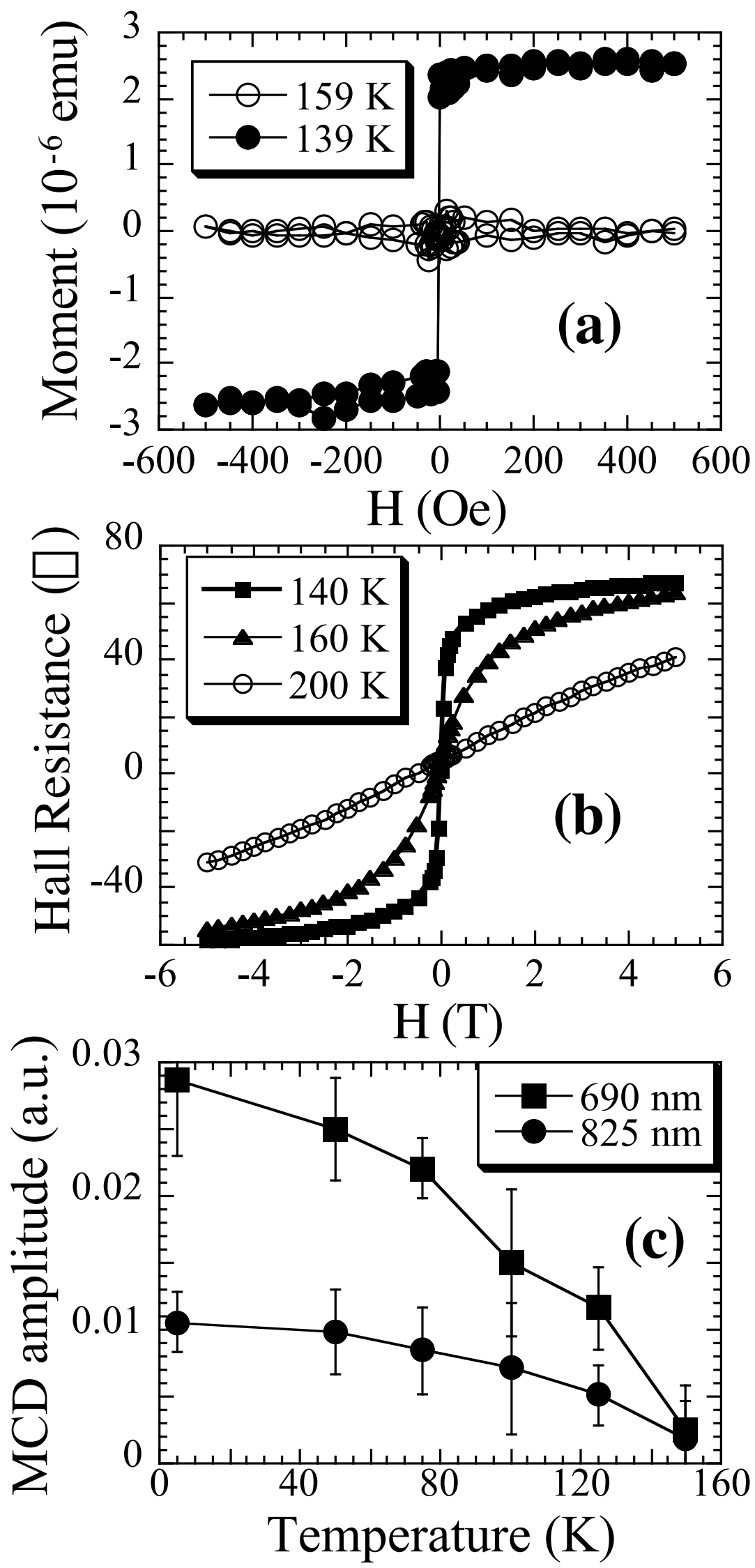

159 K
139 K
(a)
Moment (10-6 emu)
H (Oe)
140 K
160 K
200 K
(b)
Hall Resistance ( )
H (T)
690 nm
825 nm
(c)
MCD amplitude (a.u.)
Temperature (K)

Ku *et al* Fig. 2

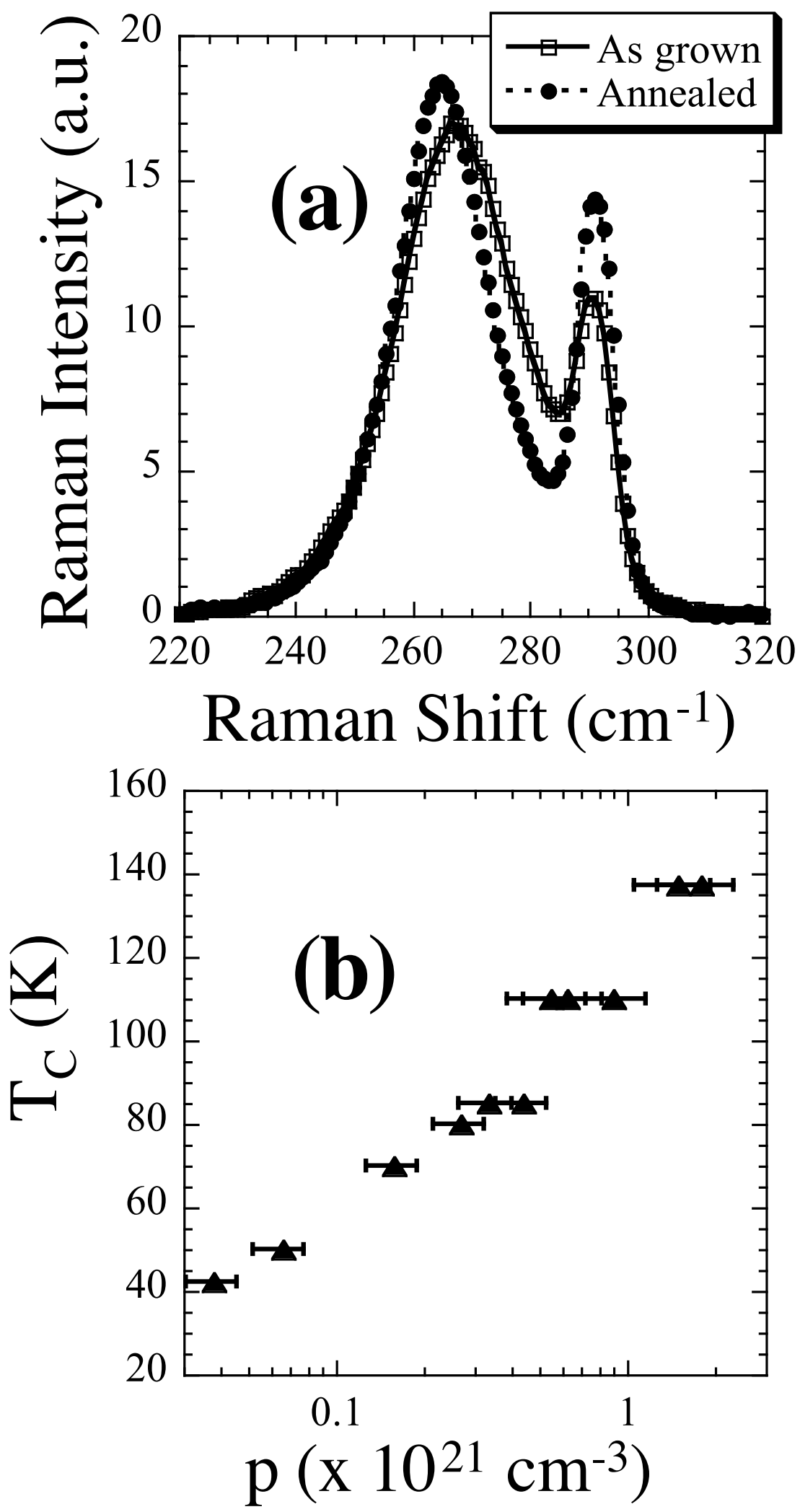

As grown
Annealed
(a)
Raman Intensity (a.u.)
Raman Shift (cm-1)
(b)
TC (K)
p (x 1021 cm-3)

Ku *et al* Fig. 3